# Polymersomes confined self-assembly: micellisation in 2D

Lorena Ruiz-Perez[1,†], Lea Messager[1,†], Jens Gaitzsch[1], Adrian Joseph[1], Ludovico Sutto[1], Francesco Luigi Gervasio[1] and Giuseppe Battaglia[1,*]

[1]Department of Chemistry, University College London, 20 Gordon Street, WC1H 0AJ, London, United Kingdom.

*Corresponding author: Prof Giuseppe Battaglia, Department of Chemistry, University College London, 20 Gordon Street, WC1H 0AJ, London, United Kingdom, tel: +44 (0)20 7679 4688 g.battaglia@ucl.ac.uk
[†]These authors have contributed equally

## Abstract

**Biological systems exploit self-assembly to create complex structures whose arrangements are finely controlled from molecular to mesoscopic level. Herein we report an example of using fully synthetic systems that mimic two levels of self-assembly. We show the formation of vesicles using amphiphilic copolymers whose chemical nature is chosen to control both membrane formation and membrane-confined interactions. We report polymersomes with patterns that emerge by engineering interfacial tension within the polymersome surface. This allows the formation of domains whose topology is tailored by the chemical synthesis paving the avenue to complex supramolecular designs functionally similar to those found in viruses and trafficking vesicles.**

Living systems are the result of a very precise and balanced hierarchical organisation of molecules and macromolecules. These are constructed with specific chemical signatures that direct supramolecular interaction between themselves and/or with water. Such interactions, typically low in energy (i.e. tens of $kT$s), allow the formation of mesoscale architectures with exquisite spatial and temporal control. This process known as self-assembly is very much ubiquitous in Nature and is at the core of any biological transformation [1]. Alongside such a positional control of molecules, Nature creates specific energy pools by enclosing chemicals into aqueous volumes using gated compartments [2]. Both compartmentalisation and positional self-assembly create structures whose surfaces express several chemistries performing their function holistically according to specific topological interactions. Biological surfaces are far from homogenous systems and organise their components according to specific (quasi)regular patterns. It is now well-established that any cell membrane has a mosaic-like structure made of dynamic nanoscale assemblies of lipids, sterols, glycols, and proteins collectively known as rafts and that these rafts control membrane signalling and trafficking [3]. Such a topological control is also conserved in smaller biological structures such as viruses, synaptic vesicles, lipoproteins and bacteria. In these, key ligands are combined into topologies with super-symmetric arrangements such as in most non-enveloped viruses[4], or have semi-ordered topologies such as in lipoproteins[5] or even into Turing-like patterns such as in most enveloped viruses [6] and endogenous trafficking vesicles [7]. Surface topology is not stochastic and is the result of an evolutionary drive often associated with a specific function. Viruses, for example, change their surface topology during maturation from a noninfectious, almost inert assembly, to an infectious cell-active structure capable of entering cells very promptly [8]. This would suggest that cellular targeting and signalling is not only controlled at a molecular level (i.e. ligand/receptor interaction) but also at a mesoscale level (i.e. how the ligand/ receptor are organised).

As our knowledge of this natural phenomenon advances so do the efforts in creating functional materials and devices that use the same principles. Among the different biomimetic efforts, polymersomes are possibly one of the few examples that encompasses both compartmentalisation and positional self-assembly into one unit. Polymersomes are vesicles formed by the self-assembly of amphiphilic block copolymers in water. In analogy to natural lipids, polymersomes can house controlled aqueous volumes to create chemical potentials across the membranes [9]. However the macromolecular nature of the polymersomes building blocks allows the design of vesicles membranes with control over their thickness, brush density, mechanical properties, and permeability [10]. Furthermore, copolymers can be designed with tuneable solubility and hence polymersomes can be made responsive to a large plethora of environmental stimuli such as pH, ionic strength, enzymatic degradation, hydrolysis, light, temperature and many others [11]. All these properties have proposed polymersomes as a very promising platform for drug and gene delivery with several examples of translation efforts of polymersomes into oncology, neurology, immunology amongst others [12]. More recently we and others have also demonstrated that polymersomes can be designed with surfaces whose topology can be controlled by polymer/ polymer interaction [13]. The mixing of two partially immiscible polymersomes-forming copolymers leads to the formation of vesicles whose surface can be patchy (binodal separation) or stripy (spinodal separation) [14]. When the two copolymers have molecular mass mismatch the same separation can causes curvature instabilities and thus the emergence of topographical features from the polymersome surface [14]. We have demonstrated that topology has a great impact on how polymersomes interact with living cells with the patchy configurations entering cells orders of magnitude more efficiently than pristine ones [15]. However whether binodal or spinodal the separation leads to a full coarsening and the formation of fully asymmetric polymersome over time [13]. Herein we propose a new approach to control the polymersomes topology using membrane confined self-assembly that creates the necessary interfacial energy to drive separation. We use molecules that can act as stabilisers decreasing the interfacial energy and hindering full phase separation. Such an entropic control over the final structure allows to translate positional self-assembly processes onto the polymersomes surface.

We synthesised three different amphiphilic block copolymers all based on poly(2-(diisopropylamino)ethyl methacrylate) (PDPA) hydrophobic block (Fig. 1). This system enables the formation of pH sensitive polymersomes that can escape the endocytic degradation once internalised by cells [16]. We combined hydrophobic PDPA with two biomedical relevant and biocompatible hydrophilic polymers: poly(2-methacryloyloxyethyl phosphorylcholine) (PMPC) and poly(ethylene oxide) (PEO) into two diblock copolymers PMPC-PDPA and PEO-PDPA and a linear triblock copolymer PEO-PDPA-PMPC. Having demonstrated that PEO-PDPA and PMPC-PDPA copolymers can form patchy and/or stripy polymersomes [15], we introduce the triblock copolymer to control the phase separation between the PEO and PMPC blocks in all effect acting as a two-dimensional surfactant (also known as line-actant [17]). In figure 1a, the structure of PMPC-PDPA is shown in addition to an optimised molecular model illustrating the spatial organisation of the copolymer hydrophilic and hydrophobic segments at their interface represented as isometric, hydrophilic-side and hydrophobic-side views. PMPC5, PEO$_{20}$ were jointed together with PDPA5 and their structure was minimised using semi-empirical method PM7 [18] with the implicit solvent model COSMO [19] and a dielectric constant of 78.4 for the hydrophilic PMPC and 4.0 for the hydrophobic PDPA. Such an analysis allows to assess how the two polymers behave at the hydrophobic/hydrophilic interface. For the PMPC-PDPA it is evident that the bulky nature of the phosphorylcholine groups of the PMPC force a larger area than that occupied by the PDPA units and indeed more than sufficient to shield PDPA from water. Conversely, in PEO-PDPA copolymers the ethylene oxide units are not sufficiently bulky to cover the hydrophobic area of PDPA (figure 1c). This mismatch in area sizes imposes the PEO to collapse onto the PDPA area to prevent its contact with water. We reported a very similar behaviour in other PEO based polymersomes using small angle x-ray scattering (SAXS) measurements [20]. Our calculations estimated, in agreement with our previous SAXS measurements [20], that at least 10 to 15 EO units are required to cover the PDPA area with a mushroom-like configuration. However, we observed that the level of confinement of the PEO within the polymersome membrane still forces the rest of the chain into a stretched configuration [20]. As for the PMPC, our model, as well as previous measurements we performed using advanced electron microscopy [21], suggests that the PMPC chains will have interchain distances that are lower than the monomer size hence a fully stretched configuration is expected [22]. Steric forces imposed by the phosphorylcholine groups also need to be taken into account in this system. Indeed the high steric forces have the capacity to hinder any chain coiling supporting our suggestion of PEO and PMPC being fully stretched. In figure 1c, we show the structure of the PEO-PDPA-PMPC triblock copolymers and using the above mentioned considerations we can estimate the triblock configuration when looped in the membrane i.e. with both PEO and PMPC facing the same side of the membrane (Figure 1d). This allows us to estimate the occupancy of the two chains forced together to calculate a two dimensional packing factor. These structural considerations are critical to understand how binary mixtures of PMPC-PDPA/PEO-PDPA-PMPC and ternary mixtures of PMPC-PDPA/PEO-PDPA-PMPC/PEO-PDPA copolymers assemble onto the polymersome surface.

We previously demonstrated that phosphotungstic acid (PTA) can be used to highlight polymers that bear carboxylic groups and by adjusting the contact time between the heavy metal and the dried polymersomes, it allows to distinguish between PMPC and PEO rich domains [15]. Fortunately, the structure of the polymersome is quite robust and it can survive controlled drying processes to allow dry state TEM. In these conditions we can visualise the polymersome surface topology with high satisfactory spatial resolution and assess the effect of copolymer compositions and their ratios. In figure 2b, we show five micrographs illustrating the effect of triblock concentration in binary systems comprising PMPC-PDPA/PEO-PDPA-PMPC triblocks. In the first one, at 10 % triblock concentration, several domains formed by the PEO chains are visible (white unstained PEO vs. the black stained PMPC). These domains have size ranging from 6 to 10nm with most of them having a circular shape and few displaying a more elongated configuration. At higher concentrations the elongated conformation becomes dominant and at triblock concentrations between 40 % and 80 % the domains merge forming a bicontinuous pattern. Finally at 90 %, the black domains seem to assume a discrete shape suggesting some sort of symmetrical arrangement. Each formulation is analysed by calculating an average spacing between the features visible on the polymersome surface and shown in Fig.2a. This analysis has been done using *ad hoc* compiled Matlab script (see SI). The resulting graph and the average spacing does not change from ca. 3 nm from 0 to 10% of triblock concentrations. This is very much similar to the dimension of a single PMPC suggesting that the triblock is either dispersed homogeneously or the emergence of potential domains is not statistically significant at this concentration. At triblock concentration of 10% we have a considerable deviation with spacing of ca. 7 nm. For higher concentrations, the spacing drops down to ca. 5 nm and stays constant for most triblock concentrations with the exception of the 100% formulation where the spacing drops down to the single PMPC chain dimension. This trend suggests that the linear triblock alone forms polymersomes with asymmetric membranes. This is with the PMPC and PEO decorating the membrane outer and inner layers respectively. Such scenario suggests that the triblock alone is more likely to adopt a bridged, as opposed to a looped conformation, across the membrane. This behaviour is well established and was reported by the Meier group [23] and we also showed similar arrangements with PMPC-PDPA-PDMEA copolymers [24].

To rationalise the shapes and pattern distribution observed experimentally on the polymersomes surface a coarse-grained model of the copolymers diffusing on a spherical surface has been devised (see supporting information for the details). Each PMPC-PDPA copolymer is represented as a single bead of mass $m_1$ representing the PMPC solvent-exposed chain. The total number of beads on the sphere surface was constant across the different simulations and resulted in N = 65539. To sample the equilibrium distribution of the beads on the sphere, the GROMACS molecular dynamics package has been used [25] to perform 300 ns-long Langevin dynamics simulations of the system at the increasing PEO-PDPA-PMPC/PDPA-PMPC ratios (5/95; 10/90; 20/80; 40/60; 50/50; 80/20; 90/10), using an inverse friction constant $t$=1 ps, an integration time step of 2 fs and the reference temperature  T = 300 K. The resulting models are displayed as see-through transparent polymersomes allowing a superposition view of the features present on both top and bottom surface areas of the sphere (Fig.2c). This is performed as an effort to reproduce the polymersomes projection images obtained from electron microscopy working in transmission mode. Indeed, we have no means to asses the opacity of the polymersome surfaces. Consequently what we observe in the imaged structures might very well be superimposed stained features from the top and base of the polymersomes. In this fashion rendering transparent the modelled polymersomes can compare them to experimentally imaged polymersomes. In figure 2c we show the corresponding results and the similarity between the simulation snapshot and the TEM micrographs is quite striking with a clear overlap of the two phases. Moreover, the shapes and pattern distribution obtained from the coarse grain model on the polymersomes surface mirror quite well the same dependence on triblock concentration as that observed experimentally. In this fashion, at low concentrations of triblock, the blue (PEO) domains emerge to form isolated circular patches, which evolve into elongated shapes as triblock concentration increases in the copolymer mixture. The elongated conformations start to merge into bicontinuous patterns, and as triblock concentration raises these patterns form a denser network. At 100% triblock concentration the dense network covers the whole surface forming the matrix with isolated orange (PMPC) domains as an inverted phase. Modelled and TEM imaged polymersomes are directly compared in Fig. 2d.

We can formalise these findings using the calculation for PMPC and PEO chain occupancy showed in figure 1f, suggesting a two-dimension micellisation process of the triblock copolymer within a PMPC diblock matrix. The PMPC-PDPA-PEO triblock has a structural configuration that does not allow a perfect packing with an area mismatch between the two hydrophilic blocks of about 0.6. Hence this area mismatch forbids the PEO chains to form regular hexagonal or triangular patterns perfectly surrounded by the PMPC chains. In figure 3 we show the arrangements using the top view of the modelled polymersomes and these display the evolution of the triblock domain formation. At low concentration, the triblock copolymers form discrete non-circular domains (we name it 2D micelles) and these domains gradually evolve into more stripe-like structures as the triblock concentration increases leading to the formation of bicontinuous surfaces. Such a process would explain the fact that the average spacing does not vary for a large range of triblock concentrations. Moreover the 2D micellisation process also suggests that we can organise the polymersome surface using the molecular design of the triblock as a building block.

The molecular dimensions of the PMPC and PEO chains within the triblock are the critical parameters for controlling the triblock self-assembly on the polymersome membrane. Indeed the results presented here propose some novel chemical design suggesting the creation of more asymmetrical configurations i.e. larger areas occupied by the PMPC chains when compared to PEO. The design of these asymmetrical configurations might very well lead to more ordered domains such those observed in viruses. However, a simpler approach to aid the self-assembly process can be achieved by adding PEO-PDPA copolymer in the system generating a ternary mixture. In figure 4a, we show the corresponding ternary diagram of the PMPC-PDPA/PEO-PDPA-PMPC/PEO-PDPA ternary polymersomes. The results are quite intriguing and albeit more complex morphologies were expected the diagram can be summarised into 4 different phases. At high concentration of triblock and low concentration of both diblocks (the top side of the diagram) we observed large domains on the polymersomes surface with some level of symmetry but generally quite disordered similar to the patterns displayed by un-controlled diblock mixtures we previously reported [14]. We appropriately named this phase as the phase-separated phase and suggest that it is the result of domains with some internal orders within disordered PEO rich areas. The scenario is more symmetrical in the rest of the diagram where three very distinct phases can be identified. At high PMPC-PDPA concentrations (left hand-side of the diagram), we observed micellar phases with PEO domains (white) dispersed within a PMPC matrix (dark grey) with an average size of about 7nm. These domains are highly convoluted and although not quite apparent, most formulations show some sort of ordering of these 2D micelles arrangements. On the other extreme of the diagram (i.e. the right hand-side) where polymersomes have a higher concentration of PEO-PDPA we observed a similar arrangement of 2D micelles, only this time the micelles seemed to be smaller (about 5.5nm). and their core were comprised of PMPC chains instead of PEO. Also this phase shows some sort of ordered arrangement of the micelles onto the surface. In the central part of the diagram, polymersomes have a surface topology with the black and white domains well connected between each other forming a bicontinuous pattern. Using the geometrical parameters calculated in figure 1 we can define the three phases. For micellar 1, the average size of 7nm corresponds to about 8 times the PDPA radius suggesting a regular hexagonal packing of 6 PEO-PDPA diblock surrounded by 12 PEO-PDPA-PMPC triblocks. For micellar 2, the domain size of 5.5nm corresponds to about 2 times the PMPC radius suggesting an hexagonal array comprising one PMPC-PDPA copolymer surrounded by 6 PEO-PDPA-PMPC triblocks. We observe that the bicontinuous phase is the arrangement showing more efficient packing in 2D and this explains why their corresponding topologies are indeed the most symmetrical ones. However, in all the three phases the presence of the PEO-PDPA copolymers clearly aids the self assembly of the triblock facilitating the formation of quite controlled patterns.

In conclusion, we demonstrated here that polymersomes can be constructed with complex surface topologies creating the necessary conditions for membrane-confined self-assembly. This is achieved by introducing in the system an interfacial energy created by the interaction of two hydrophilic blocks forced to share the same structure, the polymersome, and a triblock copolymer bearing the same two polymers that act as stabiliser. We show that this judicious copolymer mixture can self-assemble into domains with geometries and patterns that recall those formed by micelles in three dimensions. Finally, we propose more complex designs of supramolecular structures suggesting a new approach in mimicking biological units such as viruses and vesicles

**Acknowledgements**

We would like to thank Prof Steve Armes and Dr Adam Blanazs for advising us on the synthesis of the different copolymers. This work was sponsored by an ERC grant (ERC-STG-MEVIC). JG would like to thank the DFG for sponsoring his fellowship.

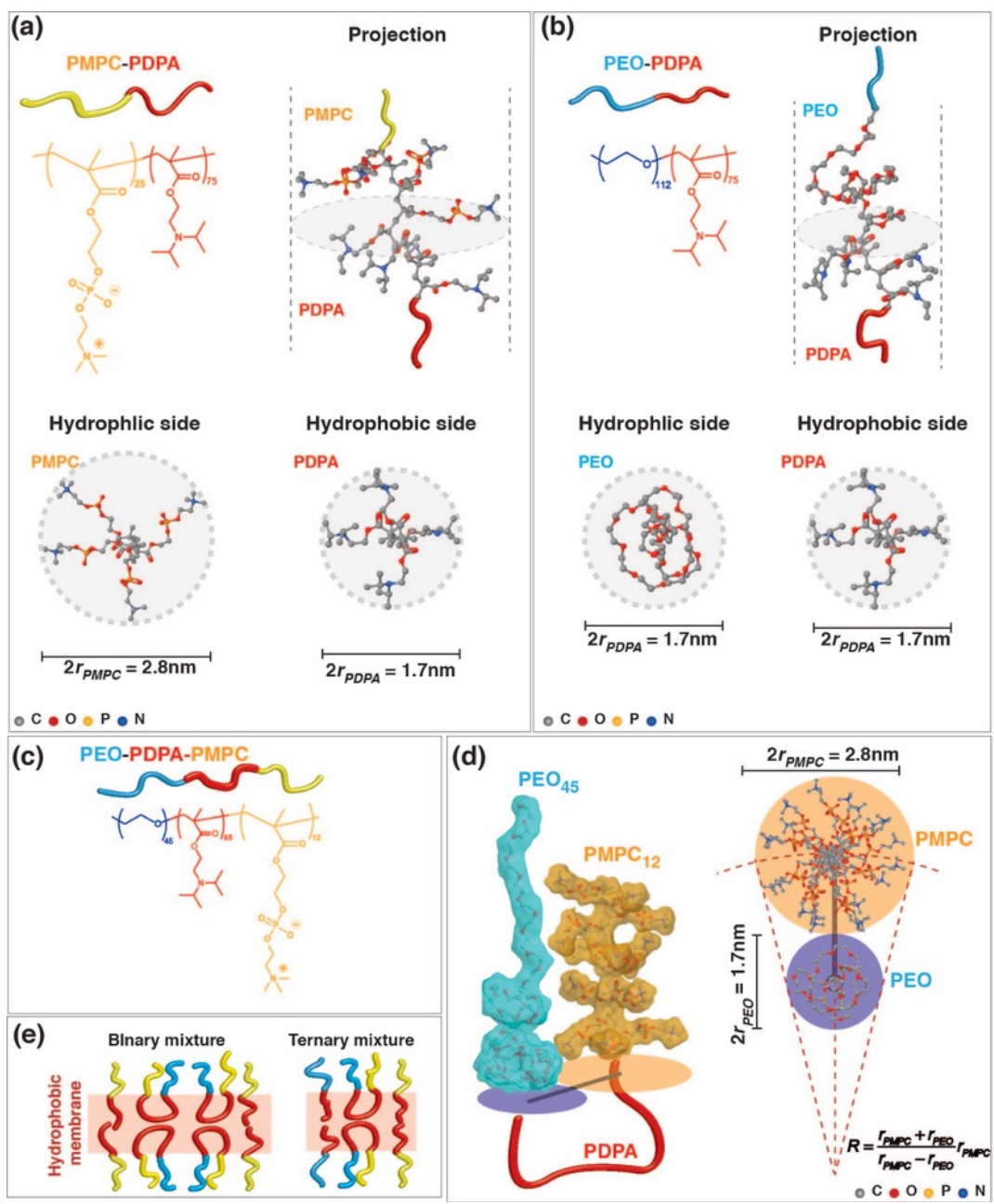

**Figure 1.** The molecular structure of PMPC-PDPA (a) and PEO-PDPA (b) with the corresponding molecular models are presented showing a possible configuration of the chains at the hydrophilic/hydrophobic interface. The models are showed as isometric projection, hydrophobic- and hydrophilic-view. The configurations were calculated using the Merck molecular force field (MMFF) algorithm. The molecular structure of the PEO-PDPA-PMPC triblock (c) and the occupancy of the two hydrophilic block PE are calculated using MMFF algorithm and represented as isometric projection and top view (d). The possible arrangements of the triblock PEO-PDPA-PMPC are shown in a binary mixture with PMPC-PDPA diblock and in a ternary mixture with PMPC-PDPA and PEO-PDPA diblocks (e).

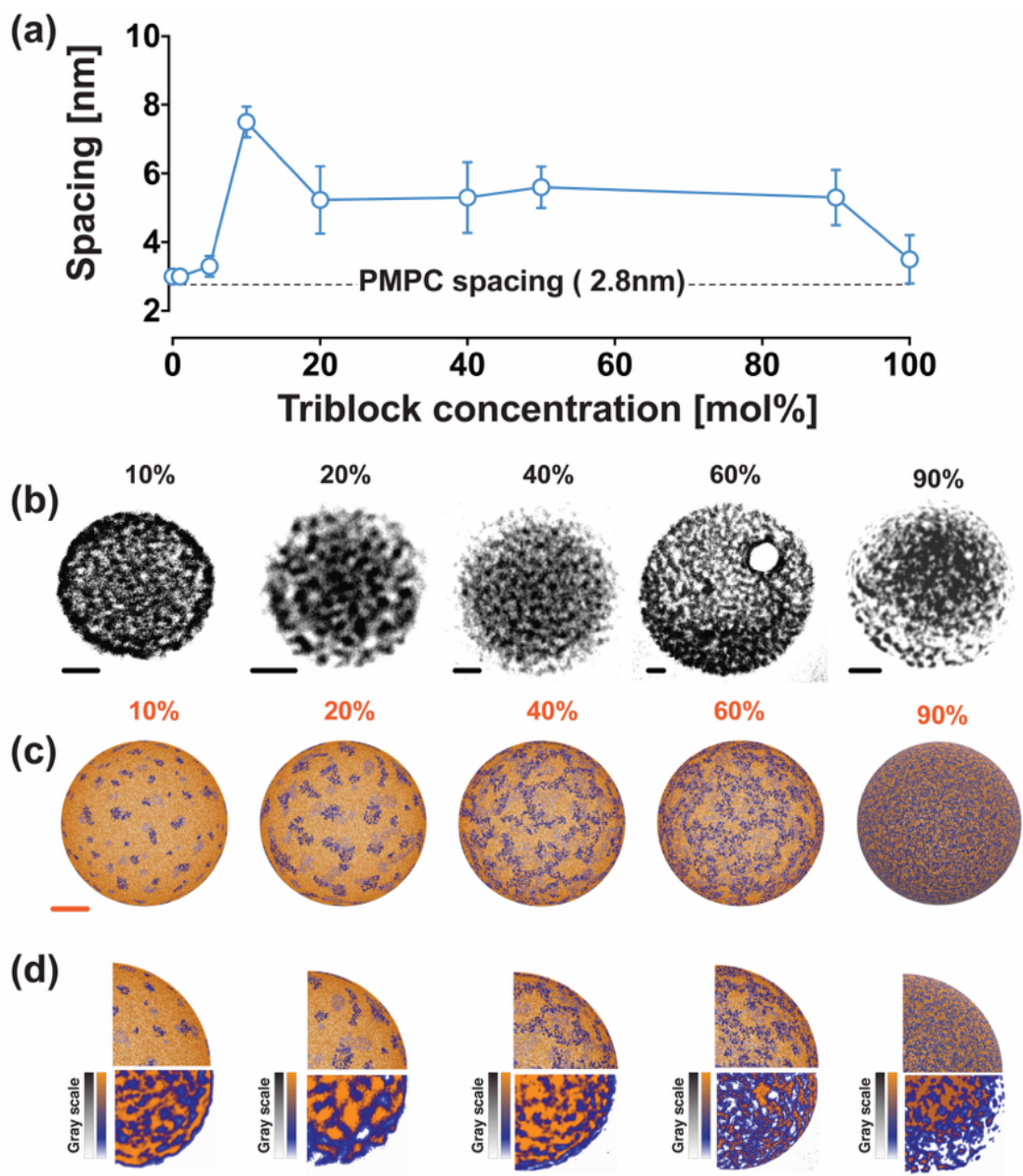

**Figure 2.** Graph showing the average spacing of the domains formed on the PMPC-PDPA/PEO-PDPA-PMPC polymersomes surace as a function of the triblock concentration. The domain spacing has been measured with MatLab. (a). TEM images (b) and coarse-grain models showed with semi-transparent top surface to simulate transmission imaging (b) of PMPC-PDPA/PEO-PDPA-PMPC polymersomes at different triblock concentrations. Comparison of the polymersome surface patterns visualised by TEM and obtained by the simulations (d)**.** Note: the TEM images are showed using a colour palette calibrated with the grayscale as showed in the figure. Scale bar is 20nm

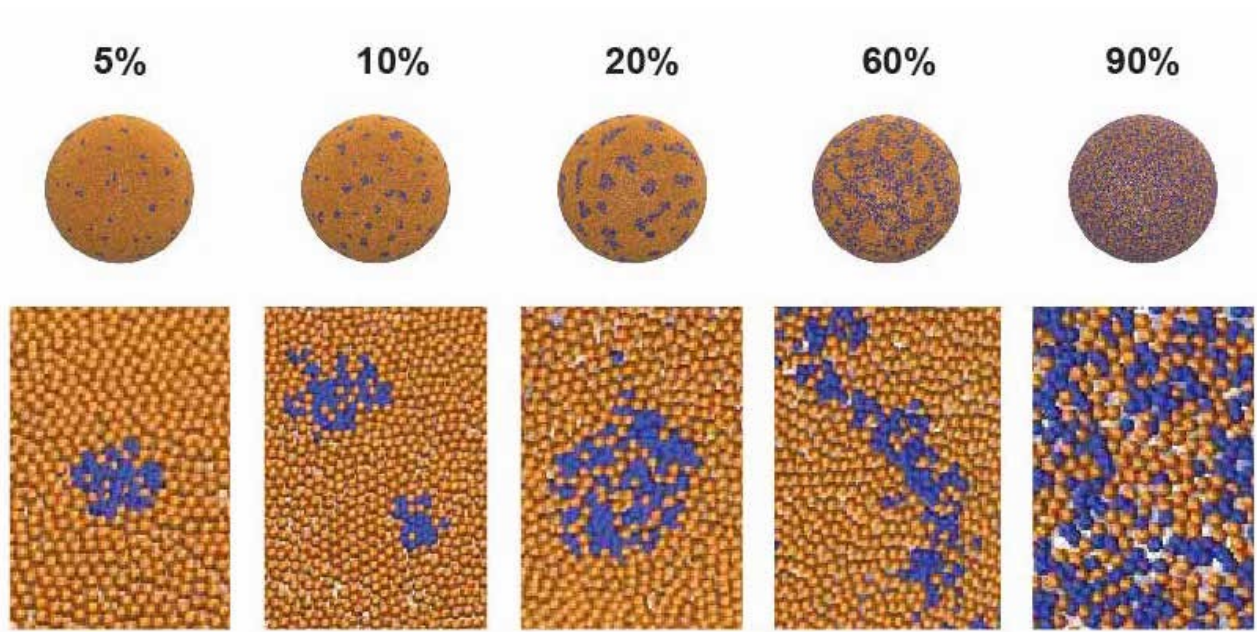

**Figure 3.** Coarse-grain simulation of PMPC-PDPA/PEO-PDPA-PMPC polymersomes at different triblock concentration displayed with a non transparent surface. The regions of interest extracted from the surface highlight our proposed mechanism of domain formation and its shape evolution as function of the triblock concentration.

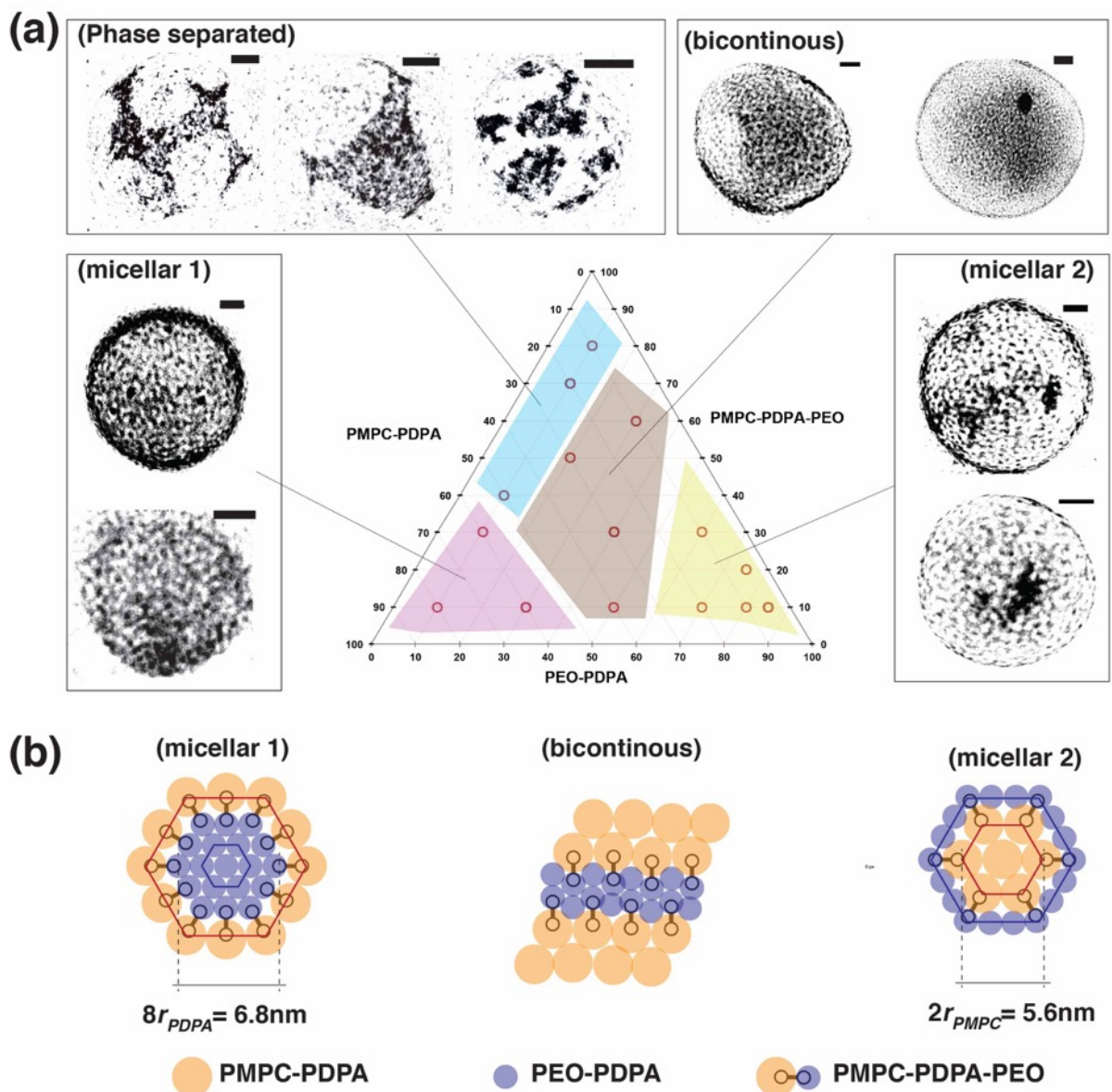

**Figure 4.** The ternary phase diagram of PMPC-PDPA/PEO-PDPA-PMPC/PEO-PDPA polymersomes **(a)**. The graphical representation of the three different phases observed in the diagram **(b).** Scale bar is 20nm

# Supporting Information –

## Polymersomes confined self-assembly: micellisation in 2D

**Lorena Ruiz-Perez**[1,*], **Lea Messager**[1,*], **Jens Gaitzsch**[1], **Adrian Joseph**[1], **Ludovico Sutto**[1], **Francesco Gervasio**[1], and **Giuseppe Battaglia**[1,*]

[1]*Department of Chemistry, Univesity College London, 20 Gordon Street, WC1H 0AJ, London, United Kingdom.*

## Materials

2-(Methacryloyloxy)ethyl phosphorylcholine monomer (MPC, 99.9% purity) was donated by Biocompatibles U.K. Ltd. Anhydrous ethanol (99 %), anhydrous methanol (≥99.8 %), 2-(diisopropylamino)ethyl methacrylate (DPA), copper(I) bromide (Cu(I)Br, 99.999%), 2,2′-bipyridine (bpy, 99%), tris(2-carboxyethyl)phosphine hydrochloride (TCEP, ≥ 98.0 %), dry triethylamine and phosphotungstic acid (PTA) were purchased from Sigma Aldrich UK. The silica gel 60 (0.063-0.200 µm) used to remove the spent ATRP catalyst CuBr was purchased from E. Merck (Darmstadt, Germany). HPLC grade dichloromethane and methanol was purchased from Fisher Scientific (Loughborough, UK). All the above were used as received.

Phosphate-buffered saline (PBS) was prepared from tablets obtained from Oxoid (Basingstoke, UK). Semi-permeable cellulose dialysis tubing (Spectra/Por 6 MWCO 1,000) was purchased from Fisher Scientific (Loughborough, UK).

### Synthesis of PMPC-PDPA

We have slightly altered a previously published procedure to synthesize linear PMPC$_{25}$-PDPA$_{70}$ diblock.[1] A solution of morpholinoethyl-bromoisobutyric acid ester (ME-Br, described previously)[1] (0.190 g, 0.00068 mol, 1 eq) was placed in a round-bottom flask before addition of MPC (5.000 g, 0.017 mol, 25 eq.). The mixture was dissolved in 5 ml of ethanol and further purged with nitrogen for 30 minutes and heated to 30 °C. Then, a mixture of bpy (0.223 g, 0.00142 mol, 2 eq.) and Cu(I)Br (0.097 g, 0.00068 mol, 1 eq.) was added under a constant nitrogen flow. The mixture was stirred for 60 minutes to yield a highly viscous brown substance and sampled for NMR to estimate conversion (gave full conversion). Meanwhile, a solution of DPA (12.27 g, 0.0576 mol, 85 eq.) in 13 mL ethanol was prepared and purged with nitrogen for 60 minutes in a separate flask. Then, the DPA solution was added to the polymerization mixture and the reaction mixture was purged for another 10 minutes and then left overnight at 30°C. After 18 h, 1H NMR analysis confirmed that the conversion was > 99 % and the reaction was opened to the atmosphere and diluted with ethanol. The solution gradually turned green, indicating oxidation of the copper-based catalyst system. The green solution was passed through silica using ethanol and evaporated partially to give an opaque solution. The solution was then dialyzed (MWCO 1,000 Da) against dichloromethane (2 times), methanol 1:1 (2 times) and water (2 times) for 8-14 hours each dialysis cycle. The polymer was first freeze-dried under vacuum and then dried at 120 °C for 2 hours in vacuum. After that the polymer was again dried for 24 hours at 90 °C, also under vacuum. (13.3 g, 77 % yield).

[1]H NMR (CDCl$_3$/MeOD - 3:1) composition: PMPC$_{25}$-PDPA$_{72}$

### Synthesis of PEO Macroinitiator

We adopted a previously published procedure by Voit *et* al. [2] Here 10g (0.002 mol) PEO$_{45}$-OH are dried in a flask at vacuum at 70 °C for 30 min. The flask is flushed with nitrogen before 20 ml dry THF is added. 0.92 g (0.004 mmol). 2-bromoisobutyric acid bromide is dissolved in 3 ml dry THF before being added to the solution. The flask is now cooled with ice and 0,303 g (0.003 mol) of dry triethylamine is added to the existing solution. The turbid mixture is stirred for 40 h at room temperature. The final macro initiator was then dialysed (MWCO = 1 kDa) against methanol (2 times) and deionised water (2 times) before being freeze-dried. Yield: 74 %.

[1]H NMR (CDCl$_3$) according to previously published ratios to PEO$_{45}$-Br. [2]

A similar procedure was used for the synthesis of PEO$_{23}$-Br macroinitiator.

### Synthesis of PEO-PDPA

We adopted a previously published procedure.[3] A solution of PEO$_{23}$-Br (0.500 g, 0.00010 mol, 1 eq) was put in a round-bottom flask and dissolved in 3 ml ethanol and DPA (2.29 g, 0.0108 mol, 20 eq.) was added. The mixture was purged with nitrogen for 30 minutes and heated to 30 °C. Then, a mixture of bpy (0.032 g, 0.00022 mol, 2 eq.) and Cu(I)Br (0.014 g, 0.00011 mol, 1 eq.) was added under a constant nitrogen flow. The mixture was stirred overnight at 30 °C to yield a highly viscous brown substance which was sampled for NMR to estimate conversion (gave full conversion). The reaction was opened to the atmosphere and diluted with ethanol. The solution gradually turned green, indicating oxidation of the copper-based catalyst system. The green solution was passed through silica using ethanol and evaporated partially to give an opaque solution. The solution was dialyzed (MWCO 1,000 Da) against dicholoromethane (2 times), methanol 1:1 (2 times) and water (2 times) for 8-14 hours each dialysis cycle. The polymer was freeze-dried under vacuum and dried at 120 °C for 2 hours in vacuum before drying another 24 hours at 90 °C, also under vacuum. (2.83 g, 98 % yield).

$^1$H NMR (CDCl$_3$) according to previously published ratios to PEO$_{23}$-PDPA$_{17}$.

### Synthesis of PEO-PDPA-PMPC

Linear triblock PEO$_{45}$-PDPA$_{60}$-PMPC$_{12}$ was synthesized by Atom Transfer Radical Polymerization (ATRP), following an adapted procedure of Blanazs and coworkers.[4] Briefly, the PEO-Br macroininiator was first synthesized according to the procedure described above. The purified compound (135 mg, 0.063 mmol, 1eq.) was subjected to vacuum for 30 min. In another round-bottom flask, DPA monomer (808 mg, 3.89 mmol, 60eq.) was diluted in ethanol to slightly decrease its viscosity and purged with nitrogen for 30 min.   The DPA solution was subsequently transferred to the PEO-Br one and the mixture was further purged with nitrogen at 30 C. Then, CuBr (9.06 mg, 0.063 mmol, 1eq.) and Bpy (19.75 mg, 0.12 mmol, 2eq.) were weighed off and added as solids in this mixture and the polymerization was carried out for at least 3 hours, until any DPA monomer could be detected by NMR. The MPC monomer (224 mg, 0.76 mmol, 12eq.) was then solubilized in ethanol and purged for 30 min with nitrogen. This solution was added to the reaction mixture and let under polymerization overnight. The highly viscous brown raw product was checked by $^1$H NMR (CDCl$_3$/MeOD - 3:1) conversion DPA 90%, MPC 99% and then opened to the atmosphere to dilute in ethanol. The solution gradually turned green, indicating oxidation of the copper-based catalyst system. The green solution was passed through silica using ethanol and evaporated partially to give an opaque solution. The solution was dialyzed (MWCO 1,000 Da) against dicholoromethane (2 times), methanol 1:1 (2 times) and water (2 times) for 8-14 hours each dialysis cycle. The polymer was freeze-dried under vacuum and dried at 120 °C for 2 hours in vacuum before drying another 24 hours at 90 °C, also under vacuum. (894 mg, 87 % yield).

The purified triblock was analysed by 1H-NMR (**Fig S1.**) and Gel Permeation Chromatography (**Fig S2.**)

¹H NMR (600 MHz, CDCl3: MetOD 3:1, ppm) δ: 0.66 (broad peak **l**, 3H, –(CH3)), 0.79 (doublet **a**, 12H, **CH₃**–CH–**CH₃**) , 1.0 (), 1.50-1.90 (broad peaks **g**, backbone), 2.42 (broad peak **b**, 2H, CH3–**CH**–CH3), 2.78 (broad peak **c**, H, –O–CH₂–**CH₂**–N–), 3.09 ( singlet f, 9H, C**H₃**–N–), 3.40 (broad peak **e**, 4H, –O–**CH₂**–**CH₂**–O–), 3.48 (broad peak **h**, 2H, –P–O –**CH₂**–**CH₂**–N–), 3.62 (broad peak **d**, 2H, –O–CH₂–**CH₂**–N–), 3.73 (broad peak **j**, 2H, –O –CH₂–**CH₂**–O–P–), 3.79, 3.94 (broad peak **i**, 2H, –P–O–**CH₂**–CH₂–N–), 4.03 (broad peak **k**, 2H, –O–**CH₂**–CH₂–O–P–), composition PEO₄₅-PDPA₆₀-PMPC₁₂.

GPC trace in 0.25% TFA aqueous solution, gives a retention volume of 8.04 mL and a PDI of 1.13

**Methods**

*Gel Permeation Chromatography* was carried out using a Malvern Viskotek GPC system (Malvern Instruments, UK) using a Novema Max 100Å Column with a Novema Max Guard Column (both PSS Polymer, Germany) with 0.25 Vol-% TFA in water as a eluent or a Resipore 100Å Column with a Resipore Guard Column (Agilent Technologies, USA) with a Chloroform/Methanol (3:1) eluent.

*NMR spectroscopy* was carried out on a Bruker AV600 spectrometer (14.1 T magnetic field strength, operating at 600 MHz for 1H NMR and 125 MHz for 13C NMR spectra).

Water was used from a TKA water purification system (Thermo Scientific, Germany)

**Patchy polymersome formation**

*Film rehydration*

Binary and ternary copolymer mixtures were prepared by mixing PMPC₂₅-PDPA₇₀ with PEO₄₅-PDPA₆₀-PMPC₁₂ and PMPC₂₅-PDPA₇₀ with PEO₄₅-PDPA₆₀ and PMPC₁₂-PEO₂₃-PDPA₁₅ at various molar ratios. Nanometer-sized polymersomes were formed by the film rehydration method. The polymer mixtures were dissolved in 2:1 v/v chloroform/methanol at 10 mg/ml total copolymer concentration in organic solvent. The solution was placed in a vacuum oven at 40°C and left overnight in order to evaporate the organic solvent. A copolymer dried thin film was formed on the sample vial surface. Rehydration of the PMPC₂₅-PDPA₇₀ /PEO₄₅-PDPA₆₀-PMPC₁₂ and PMPC₂₅-PDPA₇₀ / PEO₄₅-PDPA₆₀-PMPC₁₂ / PEO₂₃-PDPA₁₅ film was performed using 0.1 M PBS (pH 7.4) at a copolymer concentration of 5 mg/ml. The aqueous dispersions were stirred with a magnetic stirrer at 2000 rpm for two weeks at room temperature. The polymersome solution was then centrifuged 15 minutes at 500 relative centrifugal force (rcf) followed by 5 minutes at 2000 rcf using an Eppendorf Microcentrifuge. Centrifugation was performed in order to purify the solution and narrow down polymersome sizes as large and slighter particles remained in the pellet and supernatant respectively.

**Patchy polymersome characterisation**

***Transmission Electron Microscopy (TEM) imaging***

Conventional TEM imaging was performed using a FEI Tecnai G2 Spirit TEM microscope at 80 kV equipped with an Orius SC1000 camera. The polymersomes were stained using a phosphotungstic acid (PTA) solution at 0.75% (w/v). Sigma Aldrich supplied PTA at 10% (w/v) was used. The solution was prepared by dissolving 37.5 mg of PTA in boiling distilled water (5 mL). The pH was adjusted to 7.0 by adding a few drops of 5 M NaOH under continuous stirring. The PTA solution was then filtered through a 0.2 $\mu$m filter.

Copper grids were glow-discharged for 40 seconds in order to render them hydrophilic. Then 5 $\mu$L of polymersome/PBS dispersion (diluted 10-fold, concentration 0.5 mg/ mL) was deposited onto the grids for one minute. After that, the grids were blotted with filter paper and immersed into the PTA staining solution for 5s for negative staining. Then the grids were blotted again and dried under vacuum for 1 min.

***Image Analysis***

The average spacing of the domains formed on the PMPC-PDPA/PEO-PDPA-PMPC polymersome surface is calculated as a function of triblock concentration and shown on Fig.S1. In order to perform such calculations we used Matlab as it allowed for a good statistical analysis. Four different polymersomes were analysed for every prepared formulation and the number of patches used for the calculations ranged from 50-100 for each polymersome. The Matlab script stored the x and y coordinates of the domains centre of mass. The script then calculated the distances (pixels) between all the points chosen using the Pythagoras's theorem. That is, if 3 points have been chosen, namely A, B and C, Matlab calculated the distance of A from B and C, the distance of B from A and C and the distance of C from A and B. For every single set of distances Matlab will save only the shortest distance between two points. For example, if A is closer to B than it is to C, only the distance of A from B will be saved for later calculations. Final distances are converted in nm using the pixel/nm ratio previously defined. Once that all the distances have been calculated, Matlab averaged the distances and calculated the standard deviation. This method yielded results that were in very good agreement with the distances manually measured by Image J software. The resulting calculations are shown in figure S1.

As shown in Fig.2 at 10% triblock concentration, the domains are formed by the PEO chains immersed on a PMPC matrix (white unstained PEO vs. the black stained PMPC). These domains have sizes ranging from 6 to 10nm with most of them having a circular shape. This trend is observed for low triblock concentrations ranging from 1% to 10%. Accordingly the domain spacing has been measured for the white PEO domains.  At higher concentrations the elongated conformation becomes dominant and at triblock concentrations between 40% and 80% the domains merge forming a bicontinuous pattern. In these concentrations the spacing has been measured for both PEO and PMPC domains. Unsurprisingly the resulted spacing was symmetrical for both polymers.  Finally at 90%, the black PMPC domains seem to assume a discrete shape in a PEO white matrix suggesting some sort of symmetrical arrangement. For 90 % and 100% triblock concentrations the spacing have been measured for the PMPC patches. The average domain spacing as a function of triblock concentration is shown in Fig. 2

**Patchy polymersome simulation**

A coarse-grained model of the copolymers diffusing on a spherical surface has been devised. Each PMPC-PDPA copolymer is represented as a single bead of mass $m_1$ representing the PMPC solvent-exposed chain. Assuming that the PEO-PDPA-PMPC copolymer adopts a conformation where both the PEO and PMPC chains are solvent-exposed (Figure 1d), we can represent the copolymer using two connected beads of mass $m_2$ and $m_3$. To emulate the interactions of the PMPC and PEO beads, a 12-6 Lennard-Jones potential is assumed for beads of the same kind,(i.e. PMPC/PMPC and PEO/PEO), with parameters ($\mathcal{E}_1, r_1$) and ($\mathcal{E}_2, r_2$) respectively. Beads of different kinds (PMPC/PEO) repel each other with a repulsive potential

$$U(r) = \left( \frac{r_0}{r} \right)^{12}$$

where $r$ is the distance between beads. The maximum distance between the two PEO and PMPC beads representing a single PEGO-PDPA-PMPC copolymer is fixed at $d_{max}$ and it is enforced with a potential $U_R(r) = 0$ for $r < d_{max}$ and $U_R(r) = k(r-d_{max})^2$ for $r > d_{max}$. The diffusion of the beads is constrained on a polymersome surface of radius $R$. Assuming values loosely connected to the corresponding physical system due to the level of coarse-graining of the parameters such as: $m_1$ = 2313 uma, $m_2$ = 774 uma, $m_3$= 993 uma, $\mathcal{E}_1$=1 kJmol$^{-1}$, $r_1$=0.3 nm, $r_2$ = 0.2 nm, $k$=104 kJ/molnm$^{-2}$, $d_{max}$ = 1.5 nm, R=19.2 nm. The distances have been chosen to represent approximately one tenth of the observed separation distance between blocks and the beads masses to represent one tenth of the corresponding copolymer mass:

$$m_1 = \frac{1}{10}\left(25 m_{PMPC} + 74 m_{PDPA}\right)$$
$$m_2 = \frac{1}{10}\left(45 m_{PEG} + 30 m_{PDPA}\right)$$
$$m_3 = \frac{1}{10}\left(12 m_{PMPC} + 30 m_{PDPA}\right)$$

**Figures S1-S**

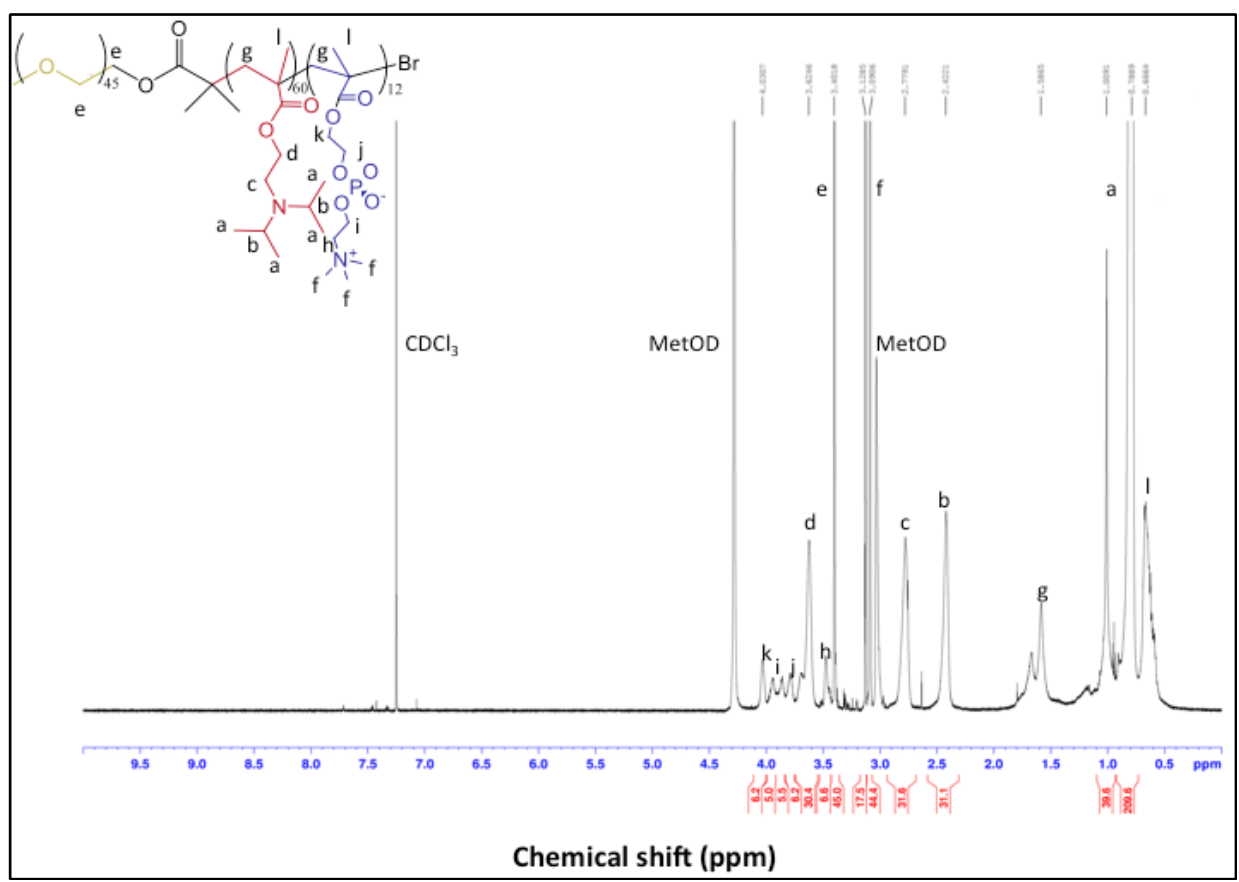

Figure S1. ¹H-NMR spectrum of PEO-PDPA-PMPC triblock copolymer in CDCl₃/MetOH (ratio 3/1); Composition: PEO₄₅-PDPA₆₀-PMPC₁₂ (relative integration of protons e versus b/c and h).

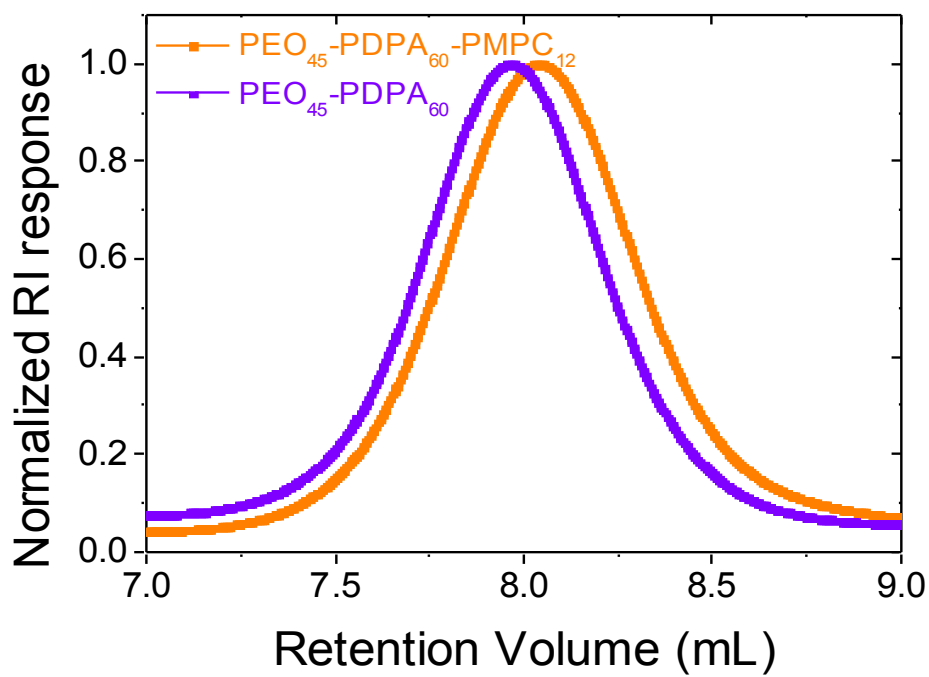

Figure S2. GPC trace of PEO-PDPA-PMPC triblock copolymer in 0.25% TFA aqueous solution, PDI =1.13 superimposed to PEO$_{45}$-PDPA$_{60}$